\documentclass[aps,reprint,twocolumn,pra,amsmath,amssymb,floatfix,
longbibliography,superscriptaddress]{revtex4-2} 

\usepackage{times}
\usepackage{physics}
\usepackage{graphicx}
\usepackage{hyperref}
\hypersetup{colorlinks=true,linkcolor=blue,citecolor=blue}
\usepackage{amsmath}
\usepackage{wrapfig}
\usepackage{siunitx}
\usepackage{bm}
\usepackage{mathtools}
\usepackage{comment}

\begin{document}

\title{
Chirality in Condensed Matter: Symmetry, Waves, and Quasiparticles
%{Chirality of condensed-matter quasiparticles: Photons, phonons, magnons and hybrid excitations}
}

\begin{abstract}
Chirality has become a recurring concept throughout condensed-matter physics, appearing in contexts ranging from optical activity and chiral phonons to magnetic excitations and hybrid quasiparticles. As its use has expanded, however, the concept has often become intertwined with related notions such as angular momentum, polarization, helicity, and nonreciprocity, obscuring its precise symmetry-based meaning. Although chirality is universally associated with broken spatial-inversion ($\mathcal{P}$) symmetry and preserved time-reversal ($\mathcal{T}$) symmetry, its physical manifestation and quantitative characterization strongly depend on the system and phenomenon under consideration. In this review, we examine chirality across photons, phonons, magnons, and hybrid excitations. We show that, while no universal measure of chirality exists, different physical systems admit different $\mathcal{P}$-odd and $\mathcal{T}$-even quantities that characterize specific chiral phenomena. We further argue that chirality is often best understood through chirality-selective interactions between waves, matter, and quasiparticles rather than as an intrinsic property of isolated excitations. From this perspective, hybridization provides a particularly promising setting for the emergence, transfer, and control of chirality in condensed-matter systems.
%\blue{Chirality has become a recurring concept throughout condensed matter physics, appearing in contexts ranging from optical activity and chiral phonons to magnetic excitations and hybrid quasiparticles. As its use has expanded, however, the concept has often become intertwined with related phenomena such as angular momentum, polarization, helicity, and nonreciprocity, obscuring its precise symmetry-based meaning. Moreover, characterstics of chirality appear to be strongly dependent on the type of system or phenomenon under consideration. In this article, we examine chirality across photons, phonons, magnons, and their hybrid excitations within the common framework based on the spatial-inversion ($\mathcal{P}$) and time-reversal ($\mathcal{T}$) symmetries. We discuss how chirality is characterized, depending on the context, by various $\mathcal{P}$-odd, $\mathcal{T}$-even quantities, and review the current understanding of chirality-selective interactions in condensed matter. We further argue that physical manifestations of chirality can often be understood only within specific interaction mechanisms that couple different fields or quasiparticles. This makes hybridization a promising arena for the emergence, transfer, and control of chirality}.  
\end{abstract}
\date{\today}

\author{Jorge Puebla}
\email{puebla.jorge.8m@kyoto-u.ac.jp}
\affiliation{Department of Electronic Science and Engineering, Kyoto University, Kyoto 615-8510, Japan}
\affiliation{{Center for Spintronics Research Network (CSRN), Kyoto University, Kyoto 611-0011, Japan}}

\author{Jun-ichiro Kishine}
\affiliation{Department of Natural Sciences, The Open University of Japan, Wakaba 2-11, Chiba 261-8586, Japan}
\affiliation{Quantum Research Center for Chirality, Institute for Molecular Science, Okazaki, Aichi 444-8585, Japan}

\author{Masashi Shiraishi}
\affiliation{Department of Electronic Science and Engineering, Kyoto University, Kyoto 615-8510, Japan}
\affiliation{{Center for Spintronics Research Network (CSRN), Kyoto University, Kyoto 611-0011, Japan}}

\author{Konstantin Y. Bliokh}
\email{konstantin.bliokh@dipc.org}
\affiliation{Donostia International Physics Center (DIPC), Donostia-San Sebasti\'an 20018, Spain}
\affiliation{{IKERBASQUE, Basque Foundation for Science, Bilbao 48009, Spain}}

\maketitle

%%%%%%%%%%%%%%%%%%%%%%%%%%%%%%%%%%%%%%%%%
\section*{Introduction}
%%%%%%%%%%%%%%%%%%%%%%%%%%%%%%%%%%%%%%%%%

Chirality is a fundamental concept that plays a profound role across multiple disciplines, spanning biology and chemistry, light-matter interactions and nanoscience, as well as high-energy physics and astrophysics \cite{Janoschek_book, Barron_book, Wagniere_book, Guijarro_book}. 
It has also become increasingly important in condensed matter physics, where it appears in optical activity \cite{Barron_book, Lindell_book}, enantiomorphic space groups of chiral crystals, recently-studied chiral phonons \cite{Juraschek2025NP, Zhang2014PRL, Zhu2018S, Ishito2023NP, Ueda2023N, Luo2023S, Choi2024NN}, as well as in topological and other surface states \cite{Hasan2010RMP, Lodahl2017Nature}, magnetic excitations, and various hybrid quasiparticles. 

Despite these developments, a general understanding of chiral properties of condensed-matter quasiparticles (beyond the well-established case of photons) is only beginning to emerge, leaving a number of unresolved questions. 
This motivates us to revisit and summarize the concept of chirality, starting from its symmetry foundations and tracing its manifestations in continuous wave fields and their quantized quasiparticle excitations.

On the one hand, chirality admits a rigorous symmetry-based definition. According to Barron \cite{Barron_book, Barron1986CPL, Barron1986CSR}, a three-dimensional object is chiral if it breaks spatial-inversion (parity, $\mathcal{P}$) symmetry while preserving time-reversal ($\mathcal{T}$) symmetry. This suggests that chirality should be associated with a {\it pseudoscalar} $\mathcal{P}$-odd and $\mathcal{T}$-even quantity. This criterion also distinguishes genuine chirality from nonreciprocal ($\mathcal{T}$-asymmetric) Faraday-like phenomena and the so-called ``false chirality'' \cite{Barron_book}. On the other hand, this symmetry-based definition does not by itself provide an explicit pseudoscalar {\it measure} of chirality. Rather, it serves as a classification criterion, indicating whether a given object is chiral without quantifying the degree or nature of its chirality.

This difficulty is fundamental: one can show that there is {\it no universal measure} of chirality \cite{Harris1999RMP, Weinberg2000CJC, Millar2005MP, Fowler2005, Pisanty2026arXiv, Berry2026arXiv}, and for any proposed chirality measure there are chiral object for which that measure vanishes.
Moreover, the chirality and handedness of an object can depend crucially on the {\it scale} and on the particular degrees of freedom under consideration. Consider, for example, a right-handed macroscopic helix assembled from left-handed chiral molecules or an optical vortex beam with a right-handed helical phase front and right-handed circular polarization. 

Therefore, the characterization of chirality is inherently context-dependent and must be tied to the specific physical phenomenon in which it manifests itself. As we show below, even the most fundamental chirality operator in relativistic field theory, or the helicity of massless particles, captures only {\it certain} aspects of chirality while failing to describe others. A consistent approach to characterizing the chirality of a physical system is therefore to start from a specific {\it chiral phenomenon} (such as circular dichroism or chirality-induced spin selectivity (CISS) \cite{Naaman2012}) and, by analyzing the underlying interaction, identify the appropriate chirality measure (of the optical field, matter, phonons, or other relevant entities) that quantifies this phenomenon \cite{Tang2010, Kishine2022IJC}.

An important consequence of the above discussion is that chirality often manifests itself not as a measurable property of an isolated constituent but through the {\it interaction} between different constituents. Accordingly, rather than asking whether an individual object is chiral, it is often more meaningful to ask whether the interaction between different objects, of the same or different physical nature, is chirality selective.

From this perspective, interactions between waves, matter, and quasiparticles, as well as the hybridization of condensed-matter excitations, provide natural settings in which chirality can emerge and become experimentally observable. A notable example is the recently developed generalized Dzyaloshinskii--Moriya (DM) interaction, which offers an effective unifying framework for describing a broad range of chiral interaction phenomena \cite{Togawa2023}.

In this review, we survey the diverse manifestations of chirality with the aim of providing a conceptual framework for chiral phenomena in condensed-matter physics. We begin by discussing the fundamental aspects of chirality in classical mechanics and relativistic field theory. 
We then examine chirality in the continuous classical wave fields underlying quantum excitations, focusing on electromagnetic waves (photons) and elastic waves (phonons). Building on this foundation, we discuss the emergence and characterization of chirality in quantized quasiparticles in crystalline solids, including phonons, magnons, and, finally, hybrid quasiparticles.

Note that the study of chirality in condensed-matter quasiparticles is still at an early stage. Even for electrons, many aspects remain incompletely understood, as exemplified by the ongoing debate surrounding CISS \cite{Naaman2012, Ray1999, Gohler2011, Bloom2024, Fransson2025}. Also, rigorous theoretical approaches to the chirality of elastic waves and phonons have emerged only recently \cite{Tateishi2025, Juraschek2025NP, Vernon2026}. In turn, the chirality of magnons and hybrid quasiparticles has only begun to be explored. We hope that the systematic progression from fundamental symmetry principles to increasingly complex systems will help clarify the physical meaning of chirality and stimulate further research into chiral phenomena in condensed-matter physics.

%%%%%%%%%%%%%%%%%%%%%%%%%%%%%%%%%%%%%%%%% 
\section*{{Fundamental aspects}}
%%%%%%%%%%%%%%%%%%%%%%%%%%%%%%%%%%%%%%%%%

\begin{figure*}[t]
    \centering 
    \includegraphics[width=0.6\textwidth]{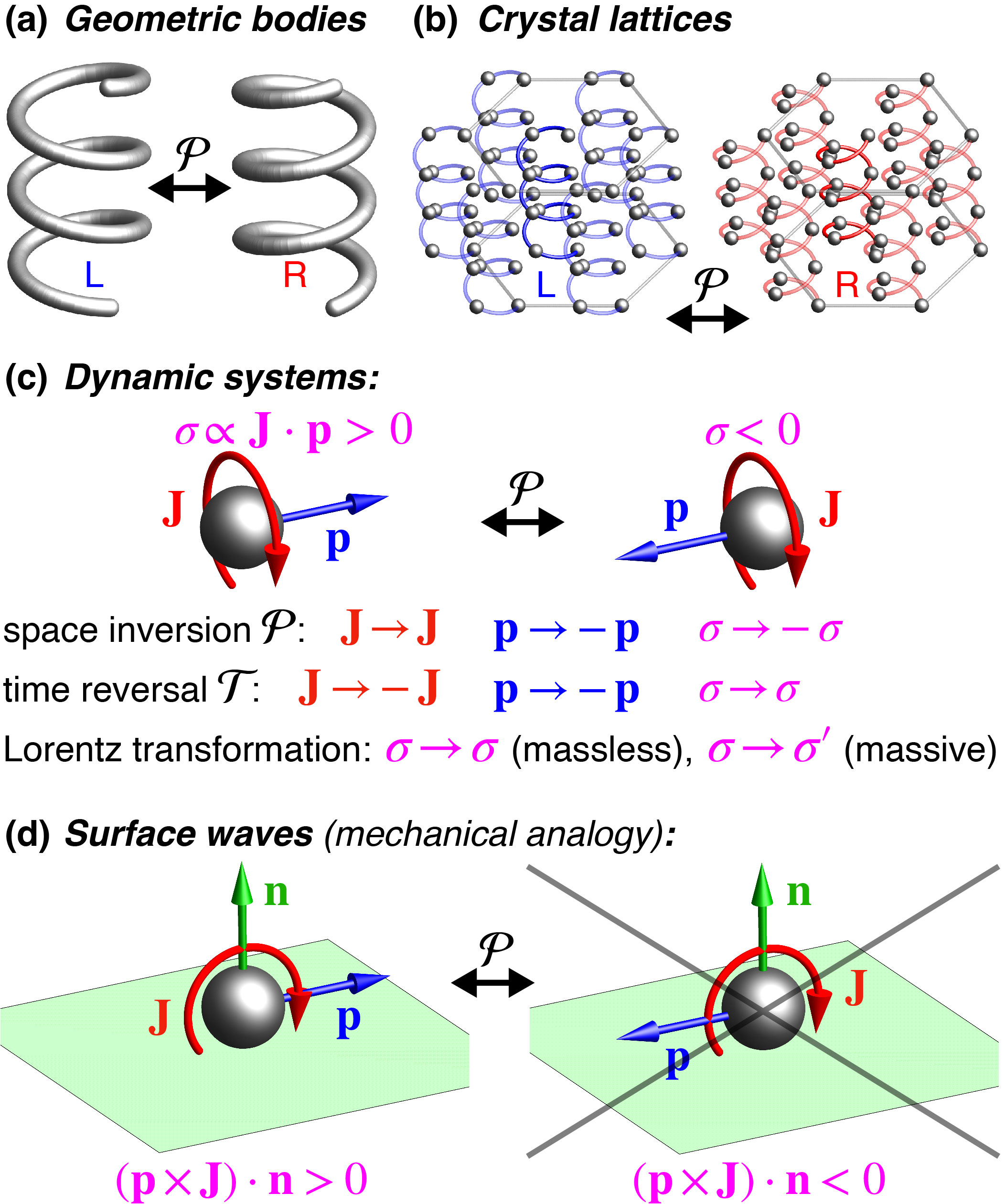} 
\caption{{\bf Basic aspects of chirality.} (a) Chiral left- and -right-handed geometric bodies that cannot be superposed with itself after the space-inversion (parity) transformation $\mathcal{P}$ and any proper rotation.
(b) Chiral crystal lattices with broken $\mathcal{P}$-symmetry, and, consequently, no improper rotational symmetry $S_n$ \cite{Sakano2020PRL}.
(c) A dynamic system, characterized by $\mathcal{P}$ and time-reversal $\mathcal{T}$ symmetries, illustrated by a classical particle carrying linear momentum ${\bf p}$ and intrinsic angular momentum (spin) ${\bf J}$. The projection of the angular momentum onto the momentum direction, i.e., helicity $\sigma = {\bf J}\cdot {\bf p}/|{\bf p}|$, is a $\mathcal{P}$-odd and $\mathcal{T}$-even pseudoscalar that naturally characterizes dynamical chirality \cite{Barron1986CPL, Barron1986CSR, Barron_book, Tang2010, Bliokh2014PRL}.
In relativistic physics, however, helicty is Lorentz-invariant only for massless particles, while the chirality of massive particles is described by the Dirac operator $\gamma^5$ \cite{Peskin_book}.
(d) Chirality of surface waves illustrated by the mechanical analogy of a ball rolling on a surface without slipping \cite{Aiello2015NP} [see also Fig.~\ref{fig:2}(c)]. The ball carries momentum ${\bf p}$ parallel to the surface, and angular momentum (spin) ${\bf J}$ directed along ${\bf n}\times {\bf p}$ (where ${\bf n}$ is the surface normal), so that $({\bf p} \times {\bf J})\cdot {\bf n} > 0$. This property of surface waves is called ``spin-momentum locking'' \cite{Hasan2010RMP, Bliokh2015Science, Lodahl2017Nature}. 
Applying the $\mathcal{P}$ transformation to the ball alone reverses the sign of $({\bf p} \times {\bf J})\cdot {\bf n}$, producing a state that is incompatible with the boundary conditions at the interface. 
Accordingly, surface-wave modes with a fixed handedness of the triad $({\bf J},{\bf p},{\bf n})$ are commonly referred to as {\it chiral}.
} 
\label{fig:1} % Adds a label for referencing the figure in the text
\end{figure*}

Chirality is a subtle concept whose precise meaning depends on 
the type of physical object under consideration and the context in which it is manifested. The earliest scientific discussion of chirality is commonly attributed to Lord Kelvin, who introduced it in 1894 as a purely geometric property of {\it static bodies}: an object is chiral if it cannot be superposed with its mirror image by any combination of translations and rotations \cite{Kelvin_book}, see Fig.~\ref{fig:1}(a). In three-dimensional (3D) space, mirror reflection is equivalent to the {spatial inversion (parity)} transformation, $\mathcal{P}: {\bf r}=(x,y,z)\rightarrow-{\bf r}$, followed by a proper rotation. Thus, chirality of a three-dimensional object implies broken $\mathcal{P}$ symmetry. (Note that in two-dimensional (2D) space, the transformation ${\bf r}=(x,y)\rightarrow-{\bf r}$ is simply a rotation by $180^\circ$, and chirality is determined by a mirror reflection, e.g., $x\rightarrow-x$.)

Chirality can also characterize {\it periodic crystal lattices}, whose symmetry is described in terms of discrete transformations and symmetries. In particular, an improper rotation, denoted by $S_n$, consists of a rotation by an angle $360^\circ/n$ ($n\in\mathbb{N}$) about a given axis followed by reflection with respect to the plane perpendicular to that axis. A crystal lattice is chiral (i.e., lacks inversion symmetry) if it possesses no $S_n$-symmetry axes. Figure~\ref{fig:1}(b) shows an example of chiral tellurium (Te) lattices, whose crystal structure belongs to the enantiomorphic (Sohncke) space groups $P3_{1}21$ and $P3_{2}21$, corresponding to right- and left-handed helical atomic chains, respectively \cite{Sakano2020PRL}.

For {\it dynamical systems}, including waves and moving bodies, spatial symmetries alone are insufficient to characterize chirality. A systematic framework distinguishing ``true'' and ``false'' chirality was developed in the 1980s by L.~D.~Barron \cite{Barron1986CPL, Barron1986CSR, Barron_book}, who recognized the central role of {time-reversal} symmetry, $\mathcal{T}$: $t\rightarrow -t$. According to Barron, a system exhibits {\it false} chirality if its $\mathcal{P}$-transformed state can be recovered by applying the $\mathcal{T}$ transformation followed by a proper rotation. Conversely, if the $\mathcal{P}$-transformed state cannot be restored in this way, the system possesses {\it true} chirality. In particular, any $\mathcal{P}$-asymmetric and $\mathcal{T}$-symmetric system is truly chiral, whereas a $\mathcal{P}$-asymmetric and $\mathcal{PT}$-symmetric system is falsely chiral.

It is instructive to examine the symmetry properties of the fundamental dynamical quantities. Linear momentum ${\bf p}$ is odd under both $\mathcal{P}$ and $\mathcal{T}$ transformations, angular momentum ${\bf J}$ is $\mathcal{P}$-even and $\mathcal{T}$-odd, whereas {\it helicity}, defined as the projection of angular momentum onto the momentum direction, ${\sigma}= {\bf J}\cdot {\bf p}/p$, is $\mathcal{P}$-odd and $\mathcal{T}$-even, see Fig.~\ref{fig:1}(c). 
Thus, helicity possesses the symmetry required of a chirality measure and can naturally serve as such for a wide class of dynamical systems.

%FFFFFFFFFFFFFFFFFFFFFFFFFFFFFFFFFFFFFFFFFFFFFFFFFFFFFFFFFFF
%FFFFFFFFFFFFFFFFFFFFFFFFFFFFFFFFFFFFFFFFFFFFFFFFFFFFFFFFFFF

However, helicity emerges in the context of {\it relativistic} quantum physics, where space and time become intertwined under Lorentz transformations between different reference frames. For massless quantum particles, such as photons, helicity does provide a Lorentz-invariant measure of chirality, as first recognized by H. Weyl \cite{Weyl1929}. In contrast, for massive particles, helicity is not Lorentz-invariant, and it can be reversed upon transformation to another frame. Therefore, a {\it frame-independent} measure of chirality is described by a distinct quantum operator: the Dirac matrix $\gamma^5$, which is $\mathcal{P}$-odd and $\mathcal{T}$-even \cite{Peskin_book}.
For massless particles, the equivalence of chirality and helicity  provides the corresponding wave equation (Weyl or Maxwell): $\gamma^5 \psi =\sigma \psi$. 
 
So far, we have considered the chirality of physical objects in free space. Classical and quantum waves {\it in media} raise an additional conceptual question: should the $\mathcal{P}$ and $\mathcal{T}$ transformations be applied to the wavefield alone or to the complete ``wave + medium'' system? Consider, for example, {\it surface waves} localized at an interface, such as surface plasmon-polaritons, surface acoustic waves, or surface electron states in topological insulators. Assuming a $\mathcal{T}$-symmetric medium, these states can be characterized by the scalar quantity $({\bf S} \times {\bf p}) \cdot {\bf n}$, where ${\bf S}$ is the spin of the wave, ${\bf p}$ is its momentum, and ${\bf n}$ is the normal to the surface \cite{Hasan2010RMP, Bliokh2015Science}. 
This quantity changes sign under inversion of either the spin or the momentum separately, while remaining invariant under their simultaneous inversion, similarly to helicity.
Consequently, if the medium is regarded as a fixed background, the surface mode breaks $\mathcal{P}$ symmetry while preserving $\mathcal{T}$ symmetry. This underlies the widespread description of such states as {\it chiral} \cite{Hasan2010RMP, Lodahl2017Nature}, see Fig.~\ref{fig:1}(d).
However, if we apply the $\mathcal{P}$ transformation to the complete ``wave + medium'' system, the surface normal ${\bf n}$ is also reversed, so that the quantity $({\bf S} \times {\bf p}) \cdot {\bf n}$ remains unchanged. Thus, surface waves under consideration are chiral {\it with respect to a fixed interface}, whereas the whole physical system is generally not. 

Finally, we mention an important manifestation of chirality in {\it wave-matter} (e.g., light-matter) {\it interactions}. In this case, the chiral properties of waves and matter naturally {\it couple} to each other. 
For example, in {\it circular dichroism}, the local chirality of a molecule couples to the helicity density of the electromagnetic field \cite{Tang2010, Bliokh2014PRL}. Likewise, isotropic chiral media (consisting of randomly oriented chiral molecules), exhibit {\it optical activity}, which results in circular dichroism (in lossy media) and, more generally, lifts the spectral degeneracy of right-handed (positive-helicity) and left-handed (negative-helicity) photons by giving them different phase velocities \cite{Lindell_book}.

The examples discussed above demonstrate that the concept of chirality depends strongly on both the physical system and the context: 3D versus 2D, static versus dynamical, non-relativistic versus relativistic, and free-space versus medium-supported systems. Therefore, the notion of chirality should always be specified for the particular problem under consideration, preferably in relation to an experimentally observable chiral phenomenon.

%%%%%%%%%%%%%%%%%%%%%%%%%%%%%%%%%%%%%%%%% 
\section*{{Continuous wavefields}}
%%%%%%%%%%%%%%%%%%%%%%%%%%%%%%%%%%%%%%%%%

Quantum particles and quasiparticles are quantized wave excitations, and their chirality appears at the level of the underlying continuous wavefields. 
Consider a real-valued 3D vector wavefield $\boldsymbol{\mathcal{F}}({\bf r},t)$. In the monochromatic (fixed-frequency) case, it can be written as $\boldsymbol{\mathcal{F}}({\bf r},t) = {\rm Re}\!\left[ {\bf F}({\bf r}) e^{-i\omega t}\right]$, where $\omega$ is the frequency, and ${\bf F}({\bf r})$ is the complex wave amplitude. 
Since the fundamental dynamical properties of a wavefield (energy, momentum, etc.) are described by its quadratic forms, the natural $\mathcal{P}$-odd and $\mathcal{T}$-even scalar quadratic form, that can characterize local chirality of the field, is:
\begin{align}
\label{helicity_F} 
\mathfrak{S} \propto \boldsymbol{\mathcal{F}} \cdot (\boldsymbol{\nabla} \times \boldsymbol{\mathcal{F}}) \rightarrow \frac{1}{2}{\rm Re}[{\bf{F}^*} \cdot (\boldsymbol{\nabla} \times {\bf{F}})]\,, 
\end{align}
where the expression on the right is the time-averaged form for a monochromatic field.
Notably, this form also universally represents wave's helicity density across various fields, including fluid mechanics and magnetohydrodynamics \cite{Woltjer1958, Moreau1961, Moffatt1969, Moffatt2014}, electromagnetism \cite{Lipkin1964, Calkin1965, Afanasiev1996, Trueba1996}, and acoustics \cite{Tateishi2025, Vernon2026}. Moreover, in a lossless isotropic medium, the total helicity (or chirality) is a {\it conserved} quantity.
Note that the chirality density is generally independent of the local spin density, given by the form ${\bf S} \propto {\rm Im} (\bf{F}^* \times \bf{F})$ \cite{Bliokh2024CP}.

%%%%%%%%%%%%%%%%%%%%%%%%%%%%%%%%%%%%%%%%%
\subsection*{Electromagnetic fields}
%%%%%%%%%%%%%%%%%%%%%%%%%%%%%%%%%%%%%%%%%

\begin{figure*}[t]
\centering 
\includegraphics[width=0.65\textwidth]{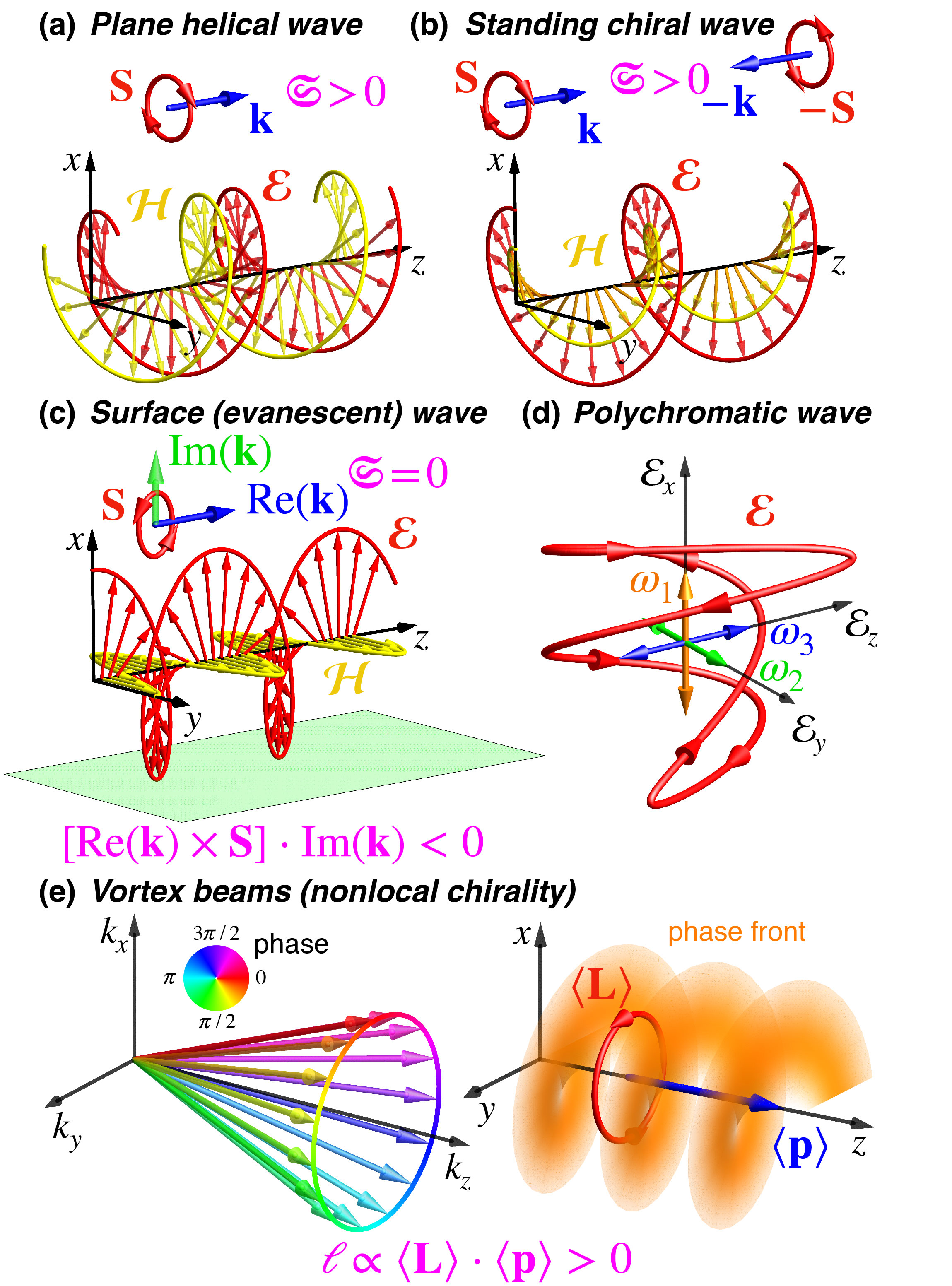} 
\caption{{\bf Examples of chiral wavefields.} (a) Right-hand circularly-polarized plane electromagnetic wave with positive helicity \eqref{helicity}. The instantaneous electric (red) and magnetic (yellow) fields are shown along the propagation $z$-axis. 
(b) A standing electromagnetic wave formed by the interference of two plane waves with opposite momenta (wavevectors) and spins. 
The electric and magnetic fields are linearly polarized at every point, but the polarization direction forms a helical distribution in space. 
Althouhg the spin and momentum densities vanish in this standing wave, the helicity density \eqref{helicity} is nonzero. 
(c) A TM-polarized surface electromagnetic wave, such as surface plasmon-polariton, has the form of an evanescent wave with the real and imaginary parts of its complex wavevector ${\bf k}$ indicating the momentum and decay directions, respectively. The helicity vanishes in this wave, whereas the triad $[{\bf S}, {\rm Im}({\bf k}), {\rm Re}({\bf k})]$ is always left-handed [cf. Fig.~\ref{fig:1}(d)]. 
(d) An example of chiral polarization trajectory of the electric field formed by the superposition of three frequency components (directed along different Cartesian axes): $\boldsymbol{\mathcal{E}} \propto [\cos(\omega_1 t +\alpha), \cos(\omega_2 t), \cos(\omega_3 t)]$, where $\omega_1=\omega_0/2$, $\omega_2=\omega_0$, $\omega_3=3\omega_0/2$, and $\alpha = \pi/6$. The corresponding cubic chirality measure \eqref{h3} is $\mathfrak{S}_{E}^{(3)} \propto \cos\alpha$ \cite{Ayuso2019NP}. 
(e) Schematics of the plane-wave spectrum and a real-space phase front of a paraxial vortex beam with topological charge $\ell=1$. It exhibits a global (nonlocal) chirality that can be quantified by the projection of the integral OAM onto the integral momentum, Eq.~\eqref{OAM}. 
All these examples have direct counterparts in elastic and other vector wavefields.}  
\label{fig:2}
\end{figure*}

From the viewpoint of relativistic field theory, photons are massless particles, for which chirality and helicity are equivalent. For a plane electromagnetic wave, the helicity is associated with the degree of circular polarization; it is $\sigma = \pm 1$ (in units of $\hbar$ per photon) for the right- and left-handed circular polarizations, respectively, see Fig.~\ref{fig:2}(a). In this case, the helicity equals the projection of the spin onto the wavevector direction: $\sigma = {\bf S}\cdot {\bf k}/k$. The chiral nature of a circularly-polarized plane wave propagating along the $z$-axis is manifested in the helical distributions of the instantaneous electric and magnetic fields, $\boldsymbol{\mathcal{E}}(z,t)$ and $\boldsymbol{\mathcal{H}}(z,t)$ [Fig.~\ref{fig:2}(a)]. However, for inhomogeneous fields, which are formed by a superposition of multiple plane waves in each point, the direct correspondence between helicity and spin (or circular polarization) no longer holds. As an example, Fig.~\ref{fig:2}(b) shows a standing electromagnetic wave produced by the interference of two counter-propagating plane waves with opposite momenta and spins but the same helicity $\sigma$. This field carries neither net momentum nor spin (it is linearly polarized in each point), but the polarization direction and instantaneous fields $\boldsymbol{\mathcal{E}}(z,t)$ and $\boldsymbol{\mathcal{H}}(z,t)$ exhibit chiral spatial distributions. 

In 1964-1965, studying conservation laws of Maxwell's equations, D. M. Lipkin \cite{Lipkin1964} and M. G. Calkin \cite{Calkin1965} introduced closely related quantities, which were later recognized as optical chirality and helicity (for monochromatic fields, they differ only by a constant factor \cite{Cameron2012}, and here we ignore this difference). These quantities correspond to distinct conservation laws and a continuous dual symmetry of Maxwell's equations, independent of the angular momentum and spatial rotations. 

Consider a monochromatic electromagnetic field, described by the complex electric and magnetic field amplitudes ${\bf E}({\bf r})$ and ${\bf H}({\bf r})$, in a homogeneous isotropic lossless nondispersive medium with permittivity $\varepsilon$ and permeability $\mu$. Introducing the ``photon  wavefunction'' $\boldsymbol{\Psi} = (\sqrt{\varepsilon} {\bf E}, \sqrt{\mu} {\bf H})/(2\sqrt{\omega})$, the helicity operator can be constructed from the canonical spin-1 and momentum operators, yielding \cite{Fernandez-Corbaton2013PRL} $\hat{\sigma} = \hat{\bf S}\cdot \hat{\bf p}/p = (c/\omega)\, \boldsymbol{\nabla} \times$, where $c=1/\sqrt{\varepsilon\mu}$ is the speed of light in the medium. This is a curl operator acting on the electric and magnetic fields. In turn, the chirality operator in this representation takes the form $\hat{\gamma}^5 = -\hat{\sigma}_2$, where $\hat{\sigma}_2$ is the Pauli matrix intertwining the electric and magnetic components of $\boldsymbol{\Psi}$. It is easy to verify that the equivalence of chirality and helicty, $\hat{\gamma}^5 \boldsymbol{\Psi} = \hat{\sigma} \boldsymbol{\Psi}$, is precisely equivalent to the monochromatic Maxwell equations. 

Calculating the local expectation values of $\hat{\sigma}$ or $\hat{\gamma}^5$, we obtain the time-averaged electromagnetic helicity or chirality density: 
\begin{align}
\label{helicity} 
\mathfrak{S}_{\rm EM} ({\bf r}) & = \frac{c}{4\omega^2} \left[ \varepsilon {\bf E}^*\cdot (\boldsymbol{\nabla} \times {\bf E}) + \mu  {\bf H}^*\cdot (\boldsymbol{\nabla} \times {\bf H})\right] \nonumber \\
& = \frac{1}{2\omega c}{\rm Im} ({\bf H}^* \cdot {\bf E})\,. 
\end{align}
This $\mathcal{P}$-odd and $\mathcal{T}$-even quantity has the universal chirality form \eqref{helicity_F}, involving both the electric and magnetic fields, and quantifies the local intrinsic ``handedness'' of the electromagnetic field. For an arbitrary superposition of circularly-polarized plane waves with the same helicity $\sigma$ [including the example in Fig.~\ref{fig:2}(b)], the fields satisfy $\sqrt{\mu}{\bf H} = - i \sigma \sqrt{\varepsilon}{\bf E}$, the local photon density is $N={\varepsilon |{\bf E}|^2}/(2\omega)$, and Eq.~\eqref{helicity} reduces to $\mathfrak{S}_{\rm EM} = \sigma N$.

Importantly, in the local dipolar interaction of light with small particles (e.g. molecules), the electromagnetic helicity density $\mathfrak{S}_{\rm EM} ({\bf r})$ couples to the chirality of the particles and governs circular dichroism \cite{Tang2010, Bliokh2014PRL}. 
Thus, this quantity allows one to engineer electromagnetic fields, including near-fields in nano-optics and plasmonics, to maximize the local chiral response \cite{Tang2010, Schaferling2012PRX}. 

Besides the chirality/helicity density characterizing circular dichroism (i.e., chiral energy absorption) of small particles, one can also define distinct field forms describing {\it chiral optical force} and {\it chiral optical torque} on the particle, allowing dynamical discrimination and sorting of opposite enantiomers \cite{Bliokh2014PRL, Canaguier-Durand2013NJP, Tkachenko2014NC, Cameron2014NJP, Kravets2019PRL, Genet2022ACSPhot, Toftul2026RMP}. In particular, the chiral force is characterized by a $\mathcal{P}$-even and $\mathcal{T}$-odd vector quantity. Furthermore, one can introduce the {\it false chirality} ($\mathcal{P}$-odd and $\mathcal{T}$-odd) of inhomogeneous electromagnetic fields. Its density is described by the quantity \cite{Bliokh2014PRL} $ {\rm Re} ({\bf H}^* \cdot {\bf E})/(2\omega c)$, also known as the ``reactive helicity'' \cite{Nieto-Vesperinas2021PRR}, which couples to the nonreciprocal magnetoelecrtic (Tellegen) response of matter \cite{Kamenetskii2023, Lindell_book, Barron_book, Bliokh2014PRL}. 

%FFFFFFFFFFFFFFFFFFFFFFFFFFFFFFFFFFFFFFFFFFFFFFFFFFFFFFFFFFF
%FFFFFFFFFFFFFFFFFFFFFFFFFFFFFFFFFFFFFFFFFFFFFFFFFFFFFFFFFFF

The helicity density \eqref{helicity} describes the field chirality in an isotropic non-chiral medium. Considering electromagnetic waves in an isotropic {\it chiral medium}, Maxwell's equations can be written as $\hat{\gamma}^5 \boldsymbol{\Psi} = (\hat{\sigma} +\kappa)\boldsymbol{\Psi}$, where $\kappa$ is the dimensionless chirality parameter of the medium \cite{Lindell_book}. This results in the splitting of the polarization-degenerate photon dispersion $\omega = kc$ into two branches $ck^\sigma = \omega(1 - \sigma \kappa)$, where $\sigma=\pm 1$ denotes the two helicity states. Thus, photons of opposite helicities propagate with different phase velocities, giving rise to the well-known phenomenon of optical activity \cite{Lindell_book}. 

Next, we comment on the chirality of surface electromagnetic waves (e.g., surface plasmon-polaritons propagating along metal-dielectric interfaces), which was briefly discussed in the previous section and illustrated in Fig.~\ref{fig:1}(d). Any surface mode has the form of an {\it evanescent wave} propagating along the surface and decaying exponentially in the direction normal to it. Such an evanescent wave can be characterized by a complex wavevector ${\bf k}$ whose real and imaginary parts describe the propagation and decay features, as shown in Fig.~\ref{fig:2}(c). Moreover, this wave exhibits elliptical polarization in the propagation plane spanned by ${\rm Re}({\bf k})$ and ${\rm Im}({\bf k})$, together with a {\it transverse spin} ${\bf S}$ orthogonal to it \cite{Bliokh2015Science, Lodahl2017Nature, Aiello2015NP, Bliokh2014NC}. Importantly, the helicity density of such an evanescent wave vanishes, $\mathfrak{S}_{EM} \equiv 0$, and its electric and magnetic field distributions are non-chiral [Fig.~\ref{fig:2}(c)]. Therefore, the chirality of this wave can only be defined with respect to the fixed interface, whose normal vector determines the handedness of the triad $({\bf S},{\bf p}, {\bf n})$ or $[{\bf S},{\rm Re}({\bf k}), {\rm Im}({\bf k})]$. Remarkably, this handedness is {\it opposite} to that expected from the mechanical rolling-ball analogy in Fig.~\ref{fig:1}(d). These properties are universal across the electromagnetic, acoustic, and other surface waves \cite{Qing2026}.

It should be emphasized that the helicity density \eqref{helicity} characterizes the electromagnetic chirality that is local in both the spatial and frequency domains. Nonlocality in either of these domains gives rise to additional manifestations of chirality. 
First, note that any monochromatic vector wavefield $\boldsymbol{\mathcal{F}}({\bf r},t) = {\rm Re}\!\left[ {\bf F}({\bf r}) e^{-i\omega t}\right]$ always traces a planar polarization ellipse in 3D space. In contrast, a polychromatic wavefield $\boldsymbol{\mathcal{F}}({\bf r},t)$ containing two or more frequency components generally traces a non-planar 3D trajectory, which itself can be chiral, i.e. $\mathcal{P}$-asymmetric \cite{Ayuso2019NP, Sugic2020PRR, Ferrer-Garcia2021PRR, Mayer2024NP}. Figure~\ref{fig:2}(d) shows an example of such a chiral polarization trajectory of the electric field $\boldsymbol{\mathcal{E}}({\bf r},t)$ at a given point ${\bf r}$, where three Cartesian field components oscillate at frequencies $\omega_1 = \omega_0/2$, $\omega_2 = \omega_0$, and $\omega_3 = \omega_1 + \omega _2 = 3\omega_0/2$. 
The chirality of this class of polarization trajectories can be characterized by the cubic form \cite{Ayuso2019NP, Pisanty2026arXiv} 
\begin{align}
\label{h3} 
\mathfrak{S}^{(3)}_{E} \propto {\rm Re}\!\left\{ {\bf E}^*(\omega_3)\cdot [{\bf E}(\omega_1) \times {\bf E}(\omega_2)] \right\}\,,
\end{align}
where ${\bf E}(\omega_i)$ are the complex amplitudes of the corresponding frequency components. This quantity, together with analogous higher-order chiral forms, governs chiral effects in local {\it nonlinear} light-matter interactions \cite{Ayuso2019NP, Mayer2024NP}.  

Second, an important example of spatially-nonlocal chiral wavefield is provided by {\it vortex beams} carrying intrinsic {\it orbital} angular momentum (OAM) \cite{Allen_book, Andrews_book, Bliokh2015PR}. 
Unlike the previously-considered chirality measures \eqref{helicity_F}--\eqref{h3}, intimately associated with the vector degrees of freedom, the chirality of vortex beams is related to the helical structure of their phase fronts. Hence, it can be described within the {\it scalar}-wave approximation (at least in the paraxial regime assumed below). 
A monochromatic vortex beam with topological charge $\ell \in \mathbb{Z}$ represents a superposition of plane waves with slightly different wavevectors whose mutual phases are described by the factor $\exp(i\ell\phi)$, with $\phi$ being the azimuthal angle in the ${\bf k}$-space, see Fig.~\ref{fig:2}(e). The resulting wavefunction has the form $\psi({\bf r}) = F(r,z) \exp(i\ell\varphi + i k_z z)$, where $(r,\varphi,z)$ are the cylindrical coordinates in real space. 
Such beams carry {\it integral} intrinsic OAM $\langle {\bf L}\rangle \simeq \ell \langle N\rangle \bar{\bf z}$ and momentum $\langle {\bf p}\rangle \simeq k_z \langle N\rangle \bar{\bf z}$, in units of $\hbar$, where $\langle N\rangle$ is the number of photons per unit propagation length, and $\bar{\bf z}$ is the unit vector of the propagation $z$-axis \cite{Allen_book, Andrews_book, Bliokh2015PR}.

The constant-phase surfaces in such beams are helical. 
However, in contrast to the quantities \eqref{helicity_F}--\eqref{h3} defined at each point in space, this form of chirality is essentially nonlocal. Similar to helicity (i.e., the projection of spin onto momentum), the global chirality of a vortex beam can be quantified by the projection of its OAM onto integral momentum: 
\begin{align}
\label{OAM} 
\mathfrak{S}_{\rm OAM} = \langle {\bf L}\rangle \cdot \frac{\langle {\bf p}\rangle}{|\langle {\bf p}\rangle|} \simeq \ell\, \langle N\rangle\,.
\end{align}
The nonlocal character of this quantity is evident from the fact that the local OAM and momentum densities (or quantum-mechanical OAM and momentum operators) satisfy ${\bf L} = {\bf r} \times {\bf p}$ and ${\bf L} \cdot {\bf p} \equiv 0$. Consequently, a local dipole interaction between vortex light and a chiral molecule is insensitive to the sign of the vortex charge $\ell$ \cite{Andrews2004OC, Araoka2005PRA, Loffler2011PRA}. Chiral light-matter interactions involving vortex beams, i.e., depending on ${\rm sgn}(\ell)$, can arise only for extended material objects interacting with the field nonlocally, starting with the electric-quadrupole interaction \cite{Brullot2016SA, Forbes2018OL, Forbes2021JPP}. However, chiral OAM-related dichroism does not appear in homogeneous (translation-invariant) chiral media \cite{Rostami2026}, where chiral effects are governed solely by polarization helicity $\sigma=\pm 1$.

It is worth noticing that although the projection \eqref{OAM} of the integral OAM onto the integral momentum provides a useful measure of the OAM-related chirality of paraxial vortex beams, an analogous construction does not apply to spin and helicity. Indeed, in the example shown in Fig.~\ref{fig:2}(b), a superposition of two waves with opposite spins and momenta but the same helicity, the total spin and momentum vanish, while helicity \eqref{helicity} does not. Moreover, for more general scalar wavefields, Eq.~\eqref{OAM} does not provide a complete characterization of chirality. For example, a superposition of two counter-propagating vortex beams with the same chirality $\ell$ (i.e., opposite OAM and momenta) produces the field $\psi({\bf r}) \propto \cos(k_z z +\ell \varphi)$, where the $z$ dependence of the beam envelope $F(r,z)$) has been neglected. This field exhibits an $\ell$-dependent chiral {\it intensity} distribution, yet the chirality measure \eqref{OAM} vanishes in this case. 

These examples, together with the different chirality measures \eqref{helicity}--\eqref{OAM}, confirm the fundamental conclusion highlighted in the Introduction: there is no universal measure of chirality \cite{Harris1999RMP, Weinberg2000CJC, Millar2005MP, Fowler2005, Pisanty2026arXiv, Berry2026arXiv}.
Rather, the chirality of a wavefield or quantum particle can be quantified only within the context of a specific physical phenomenon (e.g., its interaction with other particles) or for a restricted class of fields (e.g., electromagnetic plane waves). 
For example, in the framework of electromagnetic {\it multipoles}, it has recently been shown that the {\it electric toroidal monopole} $G_0$ possesses the required $\mathcal{P}$-odd and $\mathcal{T}$-even symmetry and, hence, provides an appropriate scalar measure for relevant chiral interactions \cite{Kishine2022IJC, Inda2024JCP}.

%%%%%%%%%%%%%%%%%%%%%%%%%%%%%%%%%%%%%%%%%
\subsection*{Elastic waves}
%%%%%%%%%%%%%%%%%%%%%%%%%%%%%%%%%%%%%%%%%

Phonons are quantized acoustic excitations in elastic media, fluids, or gases. A natural vector wavefield describing acoustic waves is the local displacement of the medium particles, $\boldsymbol{\mathcal{R}}({\bf r},t)$. Similar to the electric field in electromagnetism, in monochromatic acoustic fields, this displacement traces a polarization ellipse, which represents the actual microscopic trajectory of a medium particle. It determines the local angular momentum (spin) density of the field \cite{Bliokh2024CP, Jones1973, Nakane2018PRB, Shi2019}: ${\bf S} =\rho \boldsymbol{\mathcal{R}}\times \partial_t\boldsymbol{\mathcal{R}} \to (\rho\omega/2)\, {\rm Im} ({\bf R}^* \times {\bf R})$, where $\rho$ is the mass density of the medium.

Following the general vector-field prescription \eqref{helicity_F}, it is natural to assume that the local phonon chirality (helicity) should be described by the quadratic form $\boldsymbol{\mathcal{R}}\cdot(\boldsymbol{\nabla}\times \boldsymbol{\mathcal{R}})$ \cite{Tateishi2025}. 
Notably, this quantity vanishes identically for purely longitudinal (compression) waves because $\boldsymbol{\nabla} \times \boldsymbol{\mathcal{R}}_l \equiv {\bf 0}$. Thus, transverse (shear) fields are essential for acoustic chirality. However, the form $\boldsymbol{\mathcal{R}}\cdot(\boldsymbol{\nabla}\times \boldsymbol{\mathcal{R}})$ does not obey a conservation law in an isotropic lossless elastic medium. 

Recently, it was shown \cite{Vernon2026} that a conserved acoustic chirality (or helicity), analogous to the electromagnetic helicity \eqref{helicity}, can be constructed by introducing a second field  one can introduce the conserved acoustic chirality and helicity, similar to electromagnetic Eq.~\eqref{helicity}, by introducing a second vector field that plays the role of the magnetic field in electromagnetism. Namely, representing the transverse (shear) component of the velocity field $\boldsymbol{\mathcal{V}} =\partial_t \boldsymbol{\mathcal{R}}$ as $\boldsymbol{\mathcal{V}}_t = c_t \boldsymbol{\nabla}\times\boldsymbol{\mathcal{M}}$, where $c_t$ is the shear-wave velocity, the conserved helicity density of a monochromatic acoustic field in an isotropic elastic medium can be written as
\begin{align}
\label{helicity_acoustic}
\mathfrak{S}_A ({\bf r}) & = \frac{\rho c_t}{4} {\rm Re}\!\left[ {\bf R}^*\cdot (\boldsymbol{\nabla} \times {\bf R}) + {\bf M}^*\cdot (\boldsymbol{\nabla} \times {\bf M})\right] \nonumber \\
& = \frac{\rho\omega}{2}{\rm Im} ({\bf R}_t^* \cdot {\bf M}) + \frac{\rho\omega}{4}{\rm Im} ({\bf R}_l^* \cdot {\bf M})\,. 
\end{align}
Here the first and second terms on the right-hand side represent the purely-transverse and mixed transverse-longitudinal contributions to the acoustic helicity.  
For a longitudinal plane wave, the helicity \eqref{helicity_acoustic} vanishes, whereas for a transverse plane wave, ${\bf M} = - (c_t/\omega) {\bf k} \times {\bf R}_t$, which yields the familiar helicity relation $\mathfrak{S}_A = {\bf S}\cdot {\bf k}/k$. 
Furthermore, for an arbitrary superposition of circularly-polarized transverse plane waves with the same helicity $\sigma$, one has ${\bf M} = i \sigma {\bf R}_t$, the phonon density is $N={\rho \omega |{\bf R}_t|^2}/2$, and Eq.~\eqref{helicity_acoustic} reduces to $\mathfrak{S}_A = \sigma N$.
Thus, acoustic helicity or chirality density closely parallels its electromagnetic counterpart. It is natural to expect that it can effectively characterize local chiral interactions of phonons with other chiral particles, such as electrons or photons. 

Remarkably, all other forms of electromagnetic wavefield chirality discussed in the previous section, Eqs.~\eqref{h3}, \eqref{OAM}, and Fig.~\ref{fig:2}, have direct counterparts in acoustic wavefields. The corresponding expressions are obtained simply by replacing the electromagnetic fields with the appropriate acoustic fields. We therefore do not repeat the analogous discussion here and only emphasize that, unlike chiral light-matter interactions, chiral interactions involving phonons remain largely unexplored. At the same time, they may provide the microscopic basis for important condensed-matter phenomena, such as CISS in electron transport through chiral solids \cite{Francesco2022}.

%Remarkably, all examples of other types of chirality of electromagnetic wavefields, considered in the previous section, Eqs.~\eqref{h3} and \eqref{OAM}, and shown in Fig.~\ref{fig:2}, can be equally applied to transverse, surface, polychromatic, and vortex acoustic waves. In doing so, one only needs to substitute electromagnetic fields with the corresponding acoustic fields. Therefore, we do not repeat the same arguments, and only notice that, in contrast to light-matter interactions, chiral interactions of phonons with other particles has not been well studied so far. At the same time, they can underpin important condensed matter phenomena, such as CISS in electron transport through chiral solids.

%%%%%%%%%%%%%%%%%%%%%%%%%%%%%%%%%%%%%%%%%
\section*{Quasiparticles}
%%%%%%%%%%%%%%%%%%%%%%%%%%%%%%%%%%%%%%%%%

%%%%%%%%%%%%%%%%%%%%%%%%%%%%%%%%%%%%%%%%%
\subsection*{Phonons}
%%%%%%%%%%%%%%%%%%%%%%%%%%%%%%%%%%%%%%%%%

We now consider phonons as quantized excitations of {\it discrete} crystal lattices. For plane-wave phonons, chirality is characterized by helicity ${\bf S}\cdot {\bf k}/k$, where ${\bf S}= \sum_j M_j \boldsymbol{\mathcal{R}}_j \times \partial_t \boldsymbol{\mathcal{R}}_j$ is the spin angular momentum associated with the motion of atoms (marked by $j$) with masses $M_j$ and wave-induced displacements $\boldsymbol{\mathcal{R}}_j$ \cite{Zhang2014PRL, Juraschek2025NP, Nakane2018PRB}. As before, the helicity distinguishes right- and left-handed phonons. Operationally, a phonon is chiral when the rotational motion of the atoms propagates along the rotation axis, i.e.\ when ${\bf k}$ is not orthogonal to ${\bf S}$ \cite{Tateishi2025}. Purely in-plane rotational modes, such as those at the $K$ point of a honeycomb lattice, have ${\bf k}\cdot{\bf S}=0$: although they carry angular momentum they are not chiral and are more appropriately referred to {\it axial phonons} \cite{Juraschek2025NP, Cheong2022}.

It is important to distinguish the chirality of a phonon from the chirality of the host crystal. Left- and right-handed phonons can coexist as degenerate modes in achiral crystals. In a {\it chiral} crystal, this degeneracy is lifted throughout most of the Brillouin zone (except at the $\Gamma$ point and certain other high-symmetry points). The resulting splitting between left- and right-circularly-polarized phonon branches is the {\it acoustic activity}, originally predicted for elastic waves \cite{Portigal1968, Bhagwat1986PRB, Lakhtakia1988, Frenzel2019NC}. Microscopically, this effect can be described by {\it micropolar elasticity}, in which each point of the continuum possesses, in addition to the translational (polar) displacement $\boldsymbol{\mathcal{R}}$, an independent rotational (axial) displacement ${\bm \varphi}$. The coupling between these degrees of freedom generates a pseudoscalar contribution to the elastic energy, resulting in a helicity-dependent splitting of the transverse phonon branches, together with a roton-like minimum at finite ${\bf k}$ \cite{Kishine2020PRL}. 
Remarkably, this phonon splitting is the elastic counterpart of several analogous chiral interactions in other physical systems: optical activity, the antisymmetric spin-orbit coupling for electrons, and the DM interaction for magnons. These chiral phenomena can be unified within the framework of a {\it generalized DM effect} \cite{Togawa2023}.

A complementary description based on the discrete lattice elucidates the relevant conserved quantum numbers and provides explicit wavefunctions for chiral phonons. The screw symmetry of a chiral crystal combines a discrete rotation with a fractional translation along the helical axis. Its irreducible representations are therefore labeled by both the crystal momentum $k$ and the {\it crystal angular momentum} (CAM) $m$, which takes $n$ distinct values for an $n$-fold screw axis \cite{ZhangMurakami2022, Tsunetsugu2023, Kato2023, Tateishi2025}. For a threefold helix (line group $L3_1$, realized in Te), projection onto these irreducible representations yields symmetry-adapted eigenmodes -- a generalization of Bloch's theorem labeled by both $k$ and $m$ \cite{Tateishi2025}:
\begin{equation}
\label{eq:cam-bloch}
\begin{aligned}
{\bf e}_m(k) &= \sum_{s=+,-,z} a^s_m(k)\,{\bf e}^s_m(k)\,,\\
{\bf e}^s_m(k) &= \frac{1}{\sqrt{3\mathcal{N}}}\sum_{j=1}^{3\mathcal{N}} e^{-i(j-1)\left(kc/3+(m-m_s)\alpha\right)}\,{\bf e}^s_j\, .
\end{aligned}
\end{equation}
Here, $\alpha=2\pi/3$, $m=0,\pm1$, $\mathcal{N}$ is the number of unit cells, $c$ the lattice constant along the helical axis, whereas ${\bf e}^z_j$ and ${\bf e}^\pm_j=({\bf e}^x_j\pm i{\bf e}^y_j)/\sqrt2$ are the longitudinal and transverse circular (chiral) basis vectors, respectively, attached to the $j$-th atom. The spin indices are $m_s=+1,-1,0$ for $s=+,-,z$, respectively. 
The amplitudes $a^s_m(k)$ satisfy $\sum_s|a^s_m(k)|^2=1$ and describe the longitudinal-transverse mixing, which is a characteristic feature of chiral lattice vibrations. 

CAM is the rotational analogue of crystal momentum: it is conserved by virtue of the discrete rotational symmetry of the lattice, whereas the spin ${\bf S}$ is generally not conserved because continuous rotational symmetry is broken. Consequently, as $k$ varies from the $\Gamma$ point to the zone boundary a phonon branch can evolve continuously from near-transverse (circularly polarized) to near-longitudinal atomic motion.

The conservation of CAM leads to a clear and experimentally accessible selection rule. In a Raman scattering process, the change in photon helicity transferred to the lattice is $\sigma_i - \sigma_s \equiv \pm m \pmod{n}$, where $\sigma_{i,s}=\pm 1$ are the incident and scattered photon helicities, $m$ is the phonon CAM, the $\pm$ sign distinguishes the Stokes and anti-Stokes processes, and $n$ is the order of the screw axis ($n=3$ for the trigonal crystals discussed below) \cite{Tateishi2025}. Because the two photon helicity branches are split in a chiral crystal, circularly-polarized Raman scattering can resolve them directly and even determine the absolute handedness of the crystal. This rule has been verified experimentally in the prototypical chiral crystals $\alpha$-HgS (cinnabar) \cite{Ishito2023NP} and tellurium \cite{IshitoTe2023}. Furthermore, signatures of chiral phonons have been observed in $\alpha$-quartz using both circularly polarized Raman scattering \cite{Oishi2024} and resonant inelastic X-ray scattering \cite{Ueda2023N}.

As mentioned in the elastic-wave section above, the next frontier is the still largely unexplored chiral interaction of phonons with other quasiparticles, particularly with the spin and orbital degrees of freedom of electrons. The electron--chiral-phonon interaction vertex obeys the CAM conservation and enables the transfer of angular momentum from phonons to electrons \cite{Tateishi2025}. Moreover, both the CAM and spin of chiral phonons can be converted into electronic orbital and spin polarization \cite{TateishiCIOS2026, KatoYokoshi2026}. These mechanisms provide the microscopic basis for proposed nonreciprocal heat transport and phonon-mediated contributions to CISS.

%%%%%%%%%%%%%%%%%%%%%%%%%%%%%%%%%%%%%%%%%
\subsection*{Magnons}
%%%%%%%%%%%%%%%%%%%%%%%%%%%%%%%%%%%%%%%%%

In magnetic systems, it is useful to distinguish between {\it structural} (static) and {\it dynamical} chirality. Structural magnetic chirality refers to the handedness of an equilibrium spin texture described by the equilibrium magnetization ${\bf M}_0 ({\bf r})$. Such textures include magnetic helices and skyrmion lattices, and, like the geometric chirality of a crystal [Fig.~\ref{fig:1}(b)], their chirality originates from the absence of improper spatial symmetries \cite{Streubel2016JPD}. Two distinct mechanisms can give rise to structurally chiral magnetic states.

First, in a chiral crystal, the DM interaction locks the handedness of the magnetic helix to that of the lattice \cite{Togawa2016, Togawa2023, KishineOvchinnikov2015}.
The magnetic chirality is therefore imposed by the crystal, uniquely selected, and robust, as exemplified by monoaxial chiral helimagnets hosting a chiral soliton lattice.
Second, in the absence of the DM interaction, competing (frustrated) exchange interactions can stabilize a helical ground state of the Yoshimori type \cite{Yoshimori1959}. In this case, the handedness is selected only through spontaneous symmetry breaking, so that the two enantiomorphic helices remain energetically degenerate and generally coexist as left- and right-handed magnetic domains.

Dynamical magnetic chirality arises in magnons (spin waves), which are described by the magnetization perturbation $\mathbf{m}(\mathbf{r},t)$. Although nonreciprocity is not itself a measure of chirality, an important dynamical consequence associated with chiral magnetic order is a nonreciprocal magnon dispersion, $\omega(\mathbf{k})\neq\omega(-\mathbf{k})$. For such nonreciprocity, the simultaneous breaking of $\mathcal{P}$ and $\mathcal{T}$ is necessary but not sufficient: the dispersion remains reciprocal as long as the magnetic state retains a symmetry operation, unitary or antiunitary, that maps $\mathbf{k}$ onto $-\mathbf{k}$.

An ideal proper (zero-field) helix, with or without the DM interaction, breaks both $\mathcal{P}$ and $\mathcal{T}$, yet its magnons are reciprocal because residual symmetries still reverse $\mathbf{k}$. In the conical-helix geometry, these symmetries are removed by the appearance of a net magnetization $\mathbf{M}$ along the helical axis, and the relevant magnetochiral pseudoscalar is $\mathbf{M}\cdot\mathbf{q}$, with $\mathbf{q}$ the helical wavevector. Accordingly, in the field-induced conical state, magnons can be nonreciprocal both in DM helimagnets and infrustrated helimagnets of the Yoshimori type \cite{Yoshimori1959}, where the odd-in-$\mathbf{k}$ term $\propto\cos\theta\,[J(\mathbf{k}+\mathbf{q})-J(\mathbf{k}-\mathbf{q})]$ arises from symmetric exchange alone\cite{Cooper1962,Nagamiya1967,Kataoka1987}.

The DM interaction distinguishes the two cases in how the sign of $\mathbf{M}\cdot\mathbf{q}$ is selected and whether the nonreciprocity survives full polarization. In a chiral crystal, the DM interaction locks the handedness of the helix to the lattice, and the linear-in-$\mathbf{k}$ term it generates persists into the forced-ferromagnetic state, where both the sign and the magnitude of the magnon group velocity at $\mathbf{k}=0$ are set by the DM vector \cite{Iguchi2015,Sato2016,Seki2016}. In a Yoshimori helimagnet, the handedness is chosen by spontaneous symmetry breaking, so that degenerate left- and right-handed domains carry opposite nonreciprocity, which cancels macroscopically for equally populated domains and disappears entirely in the collinear polarized state, where inversion symmetry of the magnetic state is restored. This is the key dynamical distinction between magnetism with enforced chirality and that with spontaneously broken chirality.

As in the cases of photons and phonons, a single circularly precessing spin wave is not necessarily chiral. Such a mode carries spin angular momentum and has a well-defined sense of precession, but this sense is referenced to the equilibrium magnetization, an axial vector, and is therefore $\mathcal{P}$-even and $\mathcal{T}$-odd. For a general magnon field, a symmetry-consistent measure of chirality can instead be constructed as the pseudoscalar \eqref{helicity_F} associated with the dynamical magnetization field $\mathbf{m}(\mathbf{r},t)$:
\begin{align}
\label{helicity_m} 
\mathfrak{S}_{\bf m} \propto \mathbf{m}\cdot(\boldsymbol{\nabla}\times\mathbf{m})\,.
\end{align}
Moreover, an analogous quantity can also be defined for the {\it total} magnetization field, ${\bf M}={\bf M}_0+{\bf m}$:
$\mathfrak{S}_{\bf M} \propto \mathbf{M}\cdot(\boldsymbol{\nabla}\times\mathbf{M})$.
which naturally decomposes into structural (involving ${\bf M}_0$), dynamical (involving ${\bf m}$), and {\it mixed} contributions.

Whether these dynamical or total chirality densities satisfy a conservation law must be determined from the equations of motion. In this respect, the Landau--Lifshitz equation differs fundamentally from the electromagnetic and elastic wave equations because the equilibrium magnetization ${\bf M}_0$ breaks $\mathcal{T}$ symmetry. Consequently, the dual symmetry responsible for the conservation of optical and acoustic helicity is generally absent. In a ferromagnet, the only relevant continuous symmetry is the rotation of the spins about ${\bf M}_0$, whose associated Noether charge is the magnon number ($\mathcal{P}$-even and $\mathcal{T}$-odd) rather than chirality. Furthermore, ferromagnetic magnons possess a single sense of precession, with their spin locked to ${\bf M}_0$ rather than to the wavevector ${\bf k}$. Thus, the two-component (right-/left-handed) structure underlying helicity is absent.

This structure --- and with it a genuinely conserved magnonic chirality, analogous to helicity --- is restored in antiferromagnets. Their two-sublattice dynamics, described by a Lorentz-like $\sigma$ model, supports two degenerate counter-precessing magnon branches. In this case, the conserved chirality governs the $\mathcal{P}$-asymmetric response, including directional absorption and nonreciprocal coupling \cite{Proskurin2017PRL}. Even then, however, the two chiralities remain degenerate in a $\mathcal{P}$-symmetric antiferromagnet and become experimentally distinguishable (e.g., via a nonreciprocal dispersion, $\omega({\bf k})\neq\omega(-{\bf k})$) only when the $\mathcal{P}$ symmetry is broken by the DM interaction.

Thus, magnon chirality can be regarded as a property of the collective spin field ${\bf M}$ rather than of an individual quasiparticle. Its underlying symmetry and the corresponding measure depend on the specific magnetic model and its equations of motion. Magnon chirality becomes experimentally relevant through coupling to probes or currents with the appropriate pseudoscalar symmetry, giving rise to phenomena such as asymmetric absorption, nonreciprocal transport, and directional mode hybridization. The absence of a single universally accepted definition reflects the general principle emphasized throughout this review: chirality can only be defined within the context of a particular physical system and observable phenomenon.

%%%%%%%%%%%%%%%%%%%%%%%%%%%%%%%%%%%%%%%%%
\subsection*{Hybrid quasiparticles}
%%%%%%%%%%%%%%%%%%%%%%%%%%%%%%%%%%%%%%%%%

Hybridization provides a natural setting in which chirality can emerge, be transferred between different degrees of freedom, or become experimentally observable. Unlike isolated quasiparticles, whose chirality can often be characterized by well-defined pseudoscalars, hybrid excitations must be treated as coupled systems. Accordingly, their chirality cannot, in general, be inferred from the constituent quasiparticles alone, but must instead be determined from the symmetry of the hybrid interaction and the resulting eigenmodes.

Recent studies have shown that, when the coupling itself possesses the appropriate symmetry, hybridization can become chirality-selective and produce hybrid eigenmodes with well-defined handedness. For example, in FePSe$_3$, coherent coupling between magnons and chiral phonons gives rise to chiral magnon-polarons through symmetry-selective avoided crossings, illustrating how chirality-selective hybridization can emerge from angular-momentum and crystal-symmetry matching \cite{Cui2023NC}. More recently, it has been proposed that selective magnon-phonon coupling can itself generate genuinely chiral phonons, suggesting that hybridization may fundamentally alter the symmetry properties of the participating excitations \cite{Oppeneer2025}. 
Conversely, magnon-phonon hybridization can also transfer nonreciprocity through magnetic toroidal order, producing asymmetric propagation without necessarily implying true ($\mathcal{P}$-odd and $\mathcal{T}$-even) chirality \cite{Sukawa2026, Nomura2019}. 
These examples highlight both the richness of hybrid quasiparticles and the importance of distinguishing chirality from related, but distinct, concepts such as angular momentum, nonreciprocity, and magnetoelectric order.

In hybrid systems, where different excitations are coherently coupled, this distinction becomes particularly important. Phenomena such as nonreciprocal propagation or polarization-selectivity, commonly observed in magnon-phonon systems \cite{Liao2023PRL, Hwang2024PRL, Liao2024SA}, do not by themselves establish chirality. Rather, they reflect broken symmetries or mode asymmetries that may, but do not necessarily, produce a chiral response. A rigorous classification therefore requires analyzing the symmetry properties of the full hybrid Hamiltonian and determining whether a pseudoscalar chirality measure emerges at the level of the coupled fields. This issue becomes especially relevant near avoided crossings, where hybridization redistributes angular momentum, polarization, and parity between modes. 

A complementary perspective has recently been proposed by Trevillian and Tyberkevych  \cite{Trevillian2024}, who introduced a parity-based measure of hybridization for propagating magnon (spin-wave) modes. Rather than characterizing chirality through a pseudoscalar measure, this approach quantifies the mirror parity of the mode profile and its evolution across avoided crossings. Specifically, the parity of the $n$th propagating magnon mode is quantified through the overlap between the mode profile and its mirror-reflected ($z\to -z$) counterpart:
\begin{equation}
{P}_n = \langle m_{k,n} | \hat{{P}} | m_{k,n} \rangle
=
\int_{-L/2}^{L/2}
m_{k,n}^*(z)\, m_{k,n}(-z)\, dz ,
\end{equation}
where $m_{k,n}(z)$ is the mode profile across the film thickness $L$, and $\hat{{P}}:~z\to -z$ is the corresponding parity operator. Purely symmetric and antisymmetric modes correspond to $P_n=1$ and $P_n=-1$, respectively. In the vicinity of an avoided crossing, the propagating eigenmodes become hybridized and continuously exchange their parity character, producing coherent chiral phase-locked superpositions of symmetric and antisymmetric modes. The hybridization measure, $C_n=1-|P_n|\in[0,1]$, also quantifies the absolute value of the chirality of the hybridized modes. The upper and lower hybridized branches acquire opposite handedness.

This framework can be extended naturally to both intrinsic and synthetic antiferromagnets by redefining the parity basis in terms of the acoustic and optical magnon modes, corresponding to the in-phase and out-of-phase precession of the magnetic sublattices or coupled magnetic layers, respectively. In such systems, the parameter $P_n$ quantifies the acoustic or optical character of the propagating eigenmodes. Near an acoustic-optical mode anticrossing, the exchange of modal parity may likewise produce two hybridized branches with opposite handedness and nonreciprocal propagation. In synthetic antiferromagnets, this effect may be especially tunable because the hybridization can be controlled through the interlayer exchange, dipolar coupling, magnetic anisotropy, and magnetoelastic interactions. Whether the resulting handedness constitutes true chirality in the symmetry sense discussed throughout this review, however, remains an open question. (In a similar manner, ``chiral modes'' emerging at spectral degeneracies of non-Hermitian systems \cite{Heiss2001EPJD, Dembowski2003PRL, Peng2016PNAS} are ``chiral'' in the abstract space of eigenvectors but not necessarily chiral in real space.)

Hybridization therefore does not automatically produce chiral quasiparticles; rather, it provides a mechanism through which chirality can emerge when the coupled eigenmodes possess the appropriate symmetry. As in the cases of photons and phonons, it would be desirable to identify a corresponding pseudoscalar that quantifies the chirality of hybrid quasiparticles. This remains an open challenge. Different forms of hybridization, including magnon--magnon, magnon--phonon, magnon--photon, and photon--phonon coupling, involve different dynamical fields and therefore suggest different candidate pseudoscalars. Consequently, chirality in hybrid quasiparticles is inherently system-dependent and must be established from the symmetry properties of the full coupled system, rather than inferred solely from avoided crossings, nonreciprocity, or the angular momenta of the constituent excitations.

During the preparation of this review, our attention was drawn to a recent work \cite{Yao2026} introducing the concept of synthetic chirality, in which nonreciprocal magnon--photon hybridization arises from a nontrivial coupling phase accumulated around a closed interaction loop. The framework adopted there suggests that the chiral character of a hybrid excitation originates from the symmetry of the interaction rather than from the constituent quasiparticles considered separately. It should be emphasized, however, that the synthetic chirality introduced in Ref.~\cite{Yao2026} is defined through the symmetry of the coupling phase responsible for nonreciprocal transport, rather than through the existence of a $\mathcal{P}$-odd and $\mathcal{T}$-even chirality measure. 

%%%%%%%%%%%%%%%%%%%%%%%%%%%%%%%%%%%%%%%%%
\section*{Conclusions}
%%%%%%%%%%%%%%%%%%%%%%%%%%%%%%%%%%%%%%%%%

In this review, we have revisited the concept of chirality, from fundamental aspects to emergent manifestations in condensed-matter physics. We began with geometric bodies and dynamical systems, then reviewed diverse manifestations of chirality in continuous electromagnetic and elastic wavefields, and finally surveyed its emergence for different quasiparticles in solids: phonons, magnons, and hybrid excitations. This comparative perspective highlights both the common symmetry principles underlying chiral phenomena and the profound differences arising from the specific properties of different physical systems.

Although the criterion for chirality is well defined (broken $\mathcal{P}$ symmetry together with preserved $\mathcal{T}$ symmetry), there is no universal measure of chirality. Accordingly, we have emphasized throughout this review that different physical systems generally require different chirality measures, each tailored to a specific observable phenomenon or a particular class of objects. Furthermore, chirality is often defined operationally through chirality-selective interactions. From this perspective, a chirality measure is meaningful only insofar as it quantifies a given chiral phenomenon.

These considerations are particularly important for condensed-matter systems, where wave excitations (quasiparticles) do not exist in isolation but as constituents of a coupled ``medium + wave'' system. It is therefore essential to distinguish between the chirality of the wave itself, the chirality of the host medium, and the chirality of the wave relative to a fixed environment, as exemplified by surface-wave states.

From the materials perspective, recent demonstrations of chirality-selected crystal growth are particularly promising: the handedness inherited from an enantiopure seed can be propagated into another compound and maintained over macroscopic length scales \cite{Kousaka2022, Kousaka2023}. Such growth-based control effectively turns structural chirality into an experimentally selectable parameter, opening the possibility of systematic enantiomer-resolved studies of chiral quasiparticles and their transport and hybridization phenomena.

The study of chirality in condensed-matter quasiparticles is still in its infancy and offers numerous opportunities for future research. Important open problems include establishing physically meaningful measures of chirality for phenomena involving phonons, magnons, and hybrid quasiparticles; understanding chirality-selective interactions between different particles and quasiparticles; elucidating the role of chirality in transport phenomena such as CISS; and clarifying how chirality emerges through coupling and hybridization between different excitations. At the same time, rapid experimental advances in chiral crystals, nanophotonics, magnonics, and metamaterials continue to reveal new manifestations of chirality and new possibilities for controlling wave propagation and wave-matter interactions. We hope that the symmetry-based and phenomenon-oriented perspective presented in this review will provide a useful framework for understanding these developments and stimulate further progress in the rapidly evolving field of chiral condensed-matter physics.

\section*{Acknowledgements} 
K. Y. B. is supported by Grant PID2025-169574NB-C21 funded by MICIU/AEI/10.13039/501100011033 and is co-funded by the European Union through the project HORIZON-MSCA-2022-COFUND-01-SmartBRAIN3-101126600. J. K. acknowledges support from JSPS KAKENHI Grant Nos. 25K00962, 25H02149, 23H00091, and 26H02234, and JST ERATO Grant No. JPMJER2503, Japan. J. P. acknowledges support of JSPS KAKENHI No. 24K00576 from MEXT, Japan and from a research grant from the Murata Science and Education Foundation. 
\section*{Competing Interests}

The authors declare no competing interests.

\bibliography{refs}

\end{document}